\documentclass[10pt]{elsarticle}
\usepackage{mathrsfs}
\usepackage{amssymb}
\usepackage{amsmath,amssymb,amstext}
\usepackage{amsfonts}
\usepackage{booktabs}
\usepackage{soul}
\usepackage[figuresright]{rotating}
\usepackage{lineno,hyperref}
\usepackage[dvipsnames]{xcolor}
\usepackage{tikz}
\usepackage{algorithm,algorithmicx}
\usepackage{graphicx}
\usepackage{subcaption}
\usepackage{algpseudocode}
\usepackage{pifont}
\usepackage{mathtools}
\usepackage{multirow}
\usepackage{float }
\usepackage{array}
\usepackage{eqparbox}
\usepackage{fancyvrb}

\usepackage{lscape}
\usepackage{newlfont}
\usepackage{amsthm}

\newcommand{\etal }{\mbox{\emph{et al. }}}
 
\begin{document}

\begin{frontmatter}
\title{How efficient is contact tracing in mitigating the spread of COVID-19? A mathematical modeling approach}
\author{T. A. Biala*$^1$}
\ead{*biala.3@osu.edu}
\author{Y. O. Afolabi$^2$}
\author{A. Q. M. Khaliq$^3$}
\address{$^1$ Department of Mathematics,\\
The Ohio State University.\\
$^2$Department of Mathematics,\\
University of Louisiana at Lafayette.\\
$^3$Department of Mathematical Sciences,\\
Middle Tennessee State University, \\

}

\begin{abstract}
\quad 
Contact Tracing (CT) is one of the measures taken by government and health officials to reduce the spread of the novel coronavirus. In this paper, we investigate its efficacy by developing a compartmental model for assessing its impact on mitigating the spread of the virus. We describe the impact on the reproduction number ${\mathcal{R}_0}$ of COVID-19. In particular, we discuss the importance and relevance of parameters of the model such as the number of reported cases, effectiveness of tracking and monitoring policy, and the transmission rates to contact tracing. We describe the terms ``perfect tracking'', ``perfect monitoring'' and ``perfect reporting'' to indicate that traced contacts will be tracked while incubating, tracked contacts are efficiently monitored so that they do  not cause secondary infections, and all infected persons are reported, respectively. We consider three special scenarios: (1) perfect monitoring and perfect tracking of contacts of a reported case, (2) perfect reporting of cases and perfect monitoring of tracked reported cases  and (3) perfect reporting  and perfect tracking of contacts of reported cases. Furthermore, we gave a lower bound  on the proportion of contacts to be traced to ensure that the effective reproduction, ${\mathcal{R}_c}$, is below one and describe ${\mathcal{R}_c}$  in terms of observable quantities such as the proportion of reported and traced cases. 
%In the first scenario, we gave an upper bound on $\mathcal{R}_c$ as  $\mathcal{R}_c < \kappa\left(\dfrac{1 - s}{s}\right)$, where $\kappa$ is the average number of secondary infected traced contacts identified per untraced reported case and $s$ is the fraction of reported cases which are traced. For the second and third scenarios, we show that $\mathcal{R}_c = \kappa\left(\dfrac{1 - s}{s}\right) + \kappa_{_T}$, where $\kappa_{_T}$ is the average number of secondary infected traced contacts identified per traced reported case. 
Model simulations using the COVID-19 data obtained from John Hopkins University for some selected states in the US suggest that even late intervention of CT may reasonably reduce the transmission of COVID-19 and reduce peak hospitalizations and deaths. In particular, our findings suggest that effective monitoring policy of tracked cases and tracking of traced contacts while incubating are more crucial than tracing more contacts. The use of CT coupled with other measures such as social distancing, use of face mask, self-isolation or quarantine and {lockdowns} will greatly reduce the spread of the epidemic as well as peak hospitalizations and total deaths.
\end{abstract}
\end{frontmatter}
\section{Introduction}
Infectious diseases are often spread via direct and indirect contacts such as person-to-person contact, droplets spread, airborne transmissions and so on. Many studies \cite{Prather2020, Asadi2020, Ningthoujam2020, Galbadage2020, Somsen2020, Zhang2020b, Stadnytskyi2020, Jayaweera2020} have shown that the novel coronavirus infection spread through these means. Several measures which include social distancing, {lockdowns}, self-isolation/quarantine, use of face-masks, contact tracing amongst others,  have been enforced by authorities in reducing the spread of the virus.  In any disease outbreak therefore, contact tracing is an important tool for combating the spread of the outbreak. Contact tracing (CT) is the process whereby persons who have come in contact with an infected person are traced and monitored so that if they become symptomatic they can be efficiently isolated to reduce transmissions. Previous outbreaks of infectious diseases have been rapidly controlled with contact tracing and isolation, for example, the Ebola outbreak in West Africa in 2014, see \cite{WHO}.  Furthermore, in any disease control it is important to evaluate the efficacy of  intervention strategies such as contact
tracing. Thus, the need to explicitly measure how contact tracing can help in mitigating the transmission of coronavirus cannot be over-emphasized. A lot of studies have been conducted on the efficacy of contact tracing in relation to some diseases in the past, see \cite{Hyman2003, Klinkenberg2006, Shankar2014, Kwok2019, Vazquez2017, Guzzetta2015, Brown2015}. \\
Several mathematical models have been proposed for the dynamics of the novel coronavirus, see for example \cite{Liu2020, Wu2020, Zhao2020, Zhang2020, Biala2020, Eikenberry2020, Ngonghala2020, Khan2020, Mishra2020} and several models have incorporated contact tracing using stochastic modeling approach \cite{Kretzschmar2020} and networks \cite{Keeling2020}. However, these studies did not include the effect of contact tracing on the reproduction number of COVID-19 and the expression of this reproduction number in terms of observable quantities, a quick and efficient way of estimating the reproduction number.  In 2015, Browne et al. \cite{Brown2015} developed a deterministic model of contact tracing for Ebola epidemics which links tracing back to transmissions, and incorporates disease traits and control together with monitoring protocols. Eikenberry \etal \cite{Eikenberry2020} examined the potential of face mask use by general public to curtail the COVID-19 epidemic. Their findings suggest that face mask should be adopted nation-wide and be implemented without delay, even if most masks are homemade and of relatively low quality. Motivated essentially by the works of \cite{Brown2015} and \cite{Eikenberry2020}, the goal in this work  is to develop a deterministic model to measure the efficacy of contact tracing in mitigating the spread of COVID-19. As noted in \cite{Brown2015}, explicitly incorporating contact tracing with disease dynamics presents challenges, and population level effects of contact tracing are difficult to determine. Here, we propose a compartmental model which incorporates the disease traits and monitoring protocols. We describe the impact on the reproduction number ${\mathcal{R}_0}$ of COVID-19 and discuss the importance and relevance of parameters of the model. \\
{Anomalous and memory dependent behaviors observed in some physical phenomena have been studied using fractional-order derivative models \cite{Fomin2011, DiGiuseppe2010, Metzler2000, Caputo2003} because of their memory formalism which takes care of history and hereditary properties. This has motivated researchers to develop compartmental models using fractional-order derivatives for understanding the dynamics of disease outbreaks \cite{Diethelm2013, Sardar2015, Lu2020} . Here, we develop a time-fractional compartmental model; a modification of the SEIR model similar to the one given in \cite{Eikenberry2020} where they divided the infected population into symptomatic and asymptomatic compartments.{The time-fractional derivative is used here to account for the non-local dependence of the evolution of COVID-19 at the early stage of the outbreak. For example, at the beginning of the pandemic, much were not known about the disease, for instance, the modes of transmission of the disease were still being investigated. Also, there were still some inter-states travels across the U.S., which makes it difficult to determine exactly the number of people who may have been exposed to the disease.  These introduce non-local dependence in the evolution of COVID-19, which are easily captured using fractional derivative. This non-locality is reduced as travel restrictions and some other control measures are put in place by the governments. Moreover, the evolution of the unreported infected individuals is characterized by history effect because the time which an unreported person is initially infected is unknown exactly. This is due to the fact that the unreported individuals are in most cases asymptomatic (show no symptoms of infection), and their recovery is due to their immune systems (some recover quickly while others may take a longer time). Furthermore, unreported infected person recovers at different times after being infected, making the number of unreported infected class evolve with a memory dependency}. In the extreme scenario, our model reduces to the classical case when the fractional power approaches 1.}\

%For example, the evolution of the unreported class is characterized by a random process because the time which an unreported person is initially infected is unknown exactly. This is due to the fact that the unreported individuals are in most cases asymptomatic (show no symptoms of infection), and their recovery is due to their immune systems (some recover quickly while others may take a longer time). Furthermore, unreported infected person recovers at different times after being infected, making the number of unreported evolve with a memory dependency. In the extreme scenario our model just reduce to the classical case when the fractional power approaches 1.}\\
%{We adopt the use of fractional-derivative because the spread of an infectious disease depends not only on its current state, but also on its past state. However, integer order derivative lacks the capacity to incorporate this history dependence which by contrast is inherent in its fractional-order counterpart \cite{Podlubny1999, Diethelm2013}. Moreover, they reduce errors which may arise from neglect of parameters in the model, see table \ref{tab:Comparison}}. 
The remaining sections are organized as follows: In section 2, we begin with a discussion on the basic SEIR model and used this as a building block in deriving the new model with contact tracing. Thereafter, we consider special cases of the model and calculate their effective reproduction numbers. Furthermore, we gave a lower bound on the proportion of reported cases that must be traced to ensure the reproduction number is below one and express the reproduction number in terms of observable quantities such as average number of secondary infected persons per traced and untraced reported case.  In section 3, we perform several numerical experiments to corroborate our theoretical observations in Section 2. Finally, in section 4, we gave a comprehensive discussion on the impact of contact tracing in mitigating the spread of the virus. 
\section{Model Formulation}
\subsection{Initial Model}
We begin with a basic time-fractional SEIR model consisting of four compartments that represents  the susceptible (S), exposed (E), infected (I), recovered (R). We assumed that all the infected individuals are unreported and thus not hospitalized. The following system of differential equations models the transmission dynamics of the population:
\begin{equation*}
    \begin{split}
        _0\mathcal{D}^\alpha_{t}S(t) &= -\beta_0\dfrac{SI}{N}\\ 
        _0\mathcal{D}^\alpha_{t}E(t) &= \beta_0\dfrac{SI}{N} - \sigma E\\
       _0\mathcal{D}^\alpha_{t}I(t) &=\sigma E - \gamma I\\
       _0\mathcal{D}^\alpha_{t}R(t) &= \gamma I
    \end{split}
\end{equation*}
where {$\beta_0$} is the disease transmission rate, $\sigma$  ( $1/\sigma$) is the transition rate (disease incubation period) from the exposed class to the infectious class, $\gamma$ ($1/\gamma$) is the recovery rate (time from infectiousness until recovery) of an infected individual. We note that the parameters of the model are non-negative and  have dimensions given by  1/time$^\alpha$. This observation was originally noted in Diethelm \cite{Diethelm2013}. To alleviate this difference in dimensions, we replace the parameters  with a power $\alpha$ of new parameters to obtain the new system of equations:
\begin{equation}
\label{eqn:Inimodel}
    \begin{split}
        _0\mathcal{D}^\alpha_{t}S(t) &= -\beta_0^\alpha\dfrac{SI}{N}\\ 
        _0\mathcal{D}^\alpha_{t}E(t) &= \beta_0^\alpha\dfrac{SI}{N} - \sigma^\alpha E\\
       _0\mathcal{D}^\alpha_{t}I(t) &=\sigma^\alpha E - \gamma^\alpha I\\
       _0\mathcal{D}^\alpha_{t}R(t) &= \gamma^\alpha I.
    \end{split}
\end{equation}
%with {$\beta(t) = \beta_0^\alpha( 1- \kappa_1)\left(1 - \frac{D(t)}{N}\right)^{\kappa_2}$}.
\subsection{Preliminary Model}
The next step in the development of our model is the incorporation of hospitalized compartments (H) and splitting of the infected cases into reported (R) and unreported cases (U). This is necessary as published studies \cite{Liu2020, Chow2020, Eikenberry2020}  have shown that a considerable number of infected cases go unreported either due to  unawareness or early recovery or just perceptions of the infected individuals. We note that only the reported cases are being hospitalized during the infectious period and neglect the possibility of transmission of an hospitalized individual since they are not exposed to the general population. {Furthermore, we introduce a time-dependent transmission rate which is a function of the number of deaths (the severity of the epidemic).} Thus, we obtain the following system of time-fractional differential equations:
\begin{equation}
\label{eqn:Premodel}
    \begin{split}
        _0\mathcal{D}^\alpha_{t}S(t) &= -\beta(t)\dfrac{S}{N}\left(I_R + I_U \right)\\ 
        _0\mathcal{D}^\alpha_{t}E(t) &= \beta(t)\dfrac{S}{N}\left(I_R + I_U \right) - \sigma^\alpha E\\
       _0\mathcal{D}^\alpha_{t}I_R(t) &=\eta \sigma^\alpha E - (\gamma_{_R}^\alpha + \varphi_{_R}^\alpha) I_R\\
       _0\mathcal{D}^\alpha_{t}I_U(t) &=(1-\eta) \sigma^\alpha E - \gamma_{_U}^\alpha I_U\\
       _0\mathcal{D}^\alpha_{t}H(t) &= \varphi_{_R}^\alpha I_{R} - (\gamma_{_H}^\alpha  + \mu_{_H}^\alpha) H\\
       _0\mathcal{D}^\alpha_{t}R(t) &= \gamma_{_R}^\alpha I_R + \gamma_{_U}^\alpha I_U + \gamma_{_H}^\alpha H\\
       _0\mathcal{D}^\alpha_{t}C(t) &= \sigma^\alpha E\\
       _0\mathcal{D}^\alpha_{t}D(t) &= \mu_{_H}^\alpha H,
    \end{split}
\end{equation}
where $C(t)$ and $D(t)$ represents the total number of infected  (both reported and unreported) and the disease-induced deaths, respectively. These numbers can be explicitly calculated as
\begin{equation*}
    \begin{split}
        C(t) &= C(0) + \dfrac{\sigma^\alpha}{\Gamma(\alpha)}\int_{0}^{t}(t- s)^{\alpha-1}E(s)\,ds,\\
        D(t) &=  \dfrac{\mu_{_H}^\alpha}{\Gamma(\alpha)}\int_{0}^{t}(t- s)^{\alpha-1}H(s)\,ds.
    \end{split}
\end{equation*}
 {$\beta(t) = \beta_0^\alpha(1- \kappa_1)\left(1 - \frac{I_R + I_U + H}{N}\right)^{\kappa_2}$ is the disease transmission rate which takes into account effects of governmental actions with $\kappa_1$ being the strength of such of actions and $\kappa_2$ being {the intensity of the masses responds}.  We modelled the transmission rate as a step function of all governmental actions and the decreasing contacts among individuals responding to the proportion of the sum of infected reported, infected unreported and hospitalized}. {We note that this is a behavioral transmission rate, which incorporates the individual perception to the risk infection and governmental action strengths. This action includes all measures such as lockdowns, quarantine, hospitalization etcetera taken by the government to mitigate the spread of the disease. For more details about such behavioral transmission rates, see \cite{Capasso1978, d'Onofrio2009, He2013, Lin2020,  Buonomo2020}}. $\gamma_{_R},~\gamma_{_U},$ and $\gamma_{_H}$ are the recovery rates of a reported, unreported and hospitalized individuals, respectively. $\varphi_{_R}$  is the hospitalization rate of reported infected person and  $\mu_{_H}$ is the disease-induced death rate.
\subsection{Epidemiological parameters of the model}
For simplicity, we shall use prior studies to fix some parameters and fit the other parameters of the model. In particular, we shall fit the parameters $\beta_0, \kappa_1, \kappa_2, \mu_{_H}$ and $\alpha$ using the COVID-19 data obtained from John Hopkins University \cite{JHU} for some selected states in the US. The inclusion of $\mu_{_H}$ in the fitting parameters stems from the fact that different states have different hospitalization and death rates. Prior modeling studies suggest that the effective transmission rate $\beta_0$ ranges between 0.5-1.5 day$^{-1}$ \cite{Li2020c, Read2020, Shen2020, Eikenberry2020} and the incubation period lies in the range between 2--9 days  \cite{Furukawa2020, Biala2020, Lin2020, Wu2020}. The average of 5.1 days was estimated by Lauer \etal \cite{Lauer2020}. The infectious duration seems to have agreeing values of around 7 days for several modeling studies \cite{Li2020c, Liu2020, Eikenberry2020,Ngonghala2020, Tang2020, Fergusson2020}. Lachmann \etal \cite{Lachmann2020}  and  Li \etal \cite{Li2020d} estimated that around 88\% and 86\%, respectively, of all infections are undocumented with a 95\% credible interval. Maugeri \etal \cite{Maugeri2020} estimated that the proportion of unreported new infections by day ranged from 52.1\% to 100\%  with a total of 91.8\% of infections going unreported. Table \ref{tab:Parameters}  gives a  summary of these values  and the default values used in our model simulation.
\begin{table}[th]
    \centering
    \caption{Summary of parameter ranges and default values used in our simulation. ``Not'' denotes Notations.}
    \begin{tabular}[b]{lllll}
    \hline
      Parameters  & Not. & Ranges & References  & Default \\
      \hline
       Effective transmission rate & $\beta_0$   & 0.2--1.5 day$^{-1}$ &\cite{Fergusson2020, Li2020c, Shen2020, Eikenberry2020} & Fitted \\
      {Governmental action strength} & {$\kappa_1$} & {0.4239--0.8478} & {\cite{He2013}} & {Fitted}\\
      {Intensity of responds} &{$\kappa_2$} & {1117.3} &  {\cite{He2013}} & {Fitted}\\
       Incubation Period & $\sigma^{-1}$ &  2--14 days & \cite{Lauer2020, Lin2020, Li2020d, Eikenberry2020} & 5.1\\
       Proportion of reported new infections & $\eta$ & 0.10--0.48 & \cite{Lachmann2020, Li2020d, Maugeri2020} & 0.35\\
       Recovery rate (Reported) & $\gamma_{_R}$ & 1/14--1/3 day$^{-1}$ & \cite{Wu2020, Fergusson2020, Li2020c} & 1/7\\
       Recovery rate (Unreported) & $\gamma_{_U}$ & 1/14--1/3 day$^{-1}$ & \cite{Wu2020, Fergusson2020, Li2020c} & 1/7\\
        Recovery rate (Hospitalized) & $\gamma_{_H}$ & 1/30--1/3 day$^{-1}$ & \cite{Zhou2020, Tang2020} & 1/14\\
       Hospitalization rate & $\varphi_{_R}$ & 0.002--0.1 day$^{-1}$ & \cite{Fergusson2020, Zhou2020} & {0.05}\\
       Disease-induced death rate & $\mu_{_H}$ & 0.0001--0.1 day$^{-1}$ & \cite{Fergusson2020} & Fitted\\
       Time-fractional order & $\alpha$ & 0.5--1.0 & \cite{Biala2020} & Fitted\\
       \hline
    \end{tabular}
    
    \label{tab:Parameters}
\end{table}
\subsection{Final Model Incorporating Contact Tracing}
We incorporate CT into the preliminary model by linking the dynamics of disease model with actions of contact tracers such as monitoring and tracking. This general modeling framework is similar to a variety of CT  models employed in \cite{Hyman2003, Guzzetta2015,Brown2015}. At first, we describe the four steps of CT  for COVID-19 as described by the Center for Disease Control (CDC) \cite{CDC2020}. The Public health officer tries to identify contacts (contact investigation) by working with infected patients to help recall people they've been in contact with while being infectious. The second step (contact tracing) involves notifying and tracing of recorded contacts of the patient. Next (contact support), the officer informs and educates the contacts on the risk and dangers of being exposed. They also provide support on the next line of action for the contacts. In the case that a contact is already showing symptoms, the tracers will call an ambulance to remove/isolate the contact. Lastly (contact self-quarantine), contacts are encouraged to quarantine for a minimum of 14 days in case they also become ill.\\
To model the described process, we further make the following assumptions:
\begin{enumerate}
    \item[(a)] Only cases that are reported or hospitalized can trigger contact tracing
    \item[(b)] If a traced contact is tracked being infectious, they are immediately isolated, otherwise they are monitored for symptoms and possible isolation if symptoms develop.
    \item[(c)] We introduce parameters $\rho_{1}$ and $\rho_2$ that determine the probability or fraction of first or higher order traced contacts who will be incubating and infectious, respectively, when tracked. We simplify the model by assuming that $\rho_1 = \rho_2 = \rho$.
\end{enumerate}
Furthermore, we introduce a parameter $\beta_M$ such that $0 \leq \beta_M \leq \beta_0$ to control the efficacy of monitoring policy of contact tracers and health officers and $\epsilon$ to denote the fraction of reported cases that will be traced. With these new parameters and assumptions, we have the following system of differential equations:
\begin{equation}
\label{eqn:CTmodel}
    \begin{split}
        _0\mathcal{D}^\alpha_{t}S(t) &= -\beta(t)\dfrac{S}{N}\left(I_R + I_U \right) - \beta(t)\dfrac{SI_T}{N} - \beta_M^\alpha \dfrac{SI_M}{N}\\
        _0\mathcal{D}^\alpha_{t}E(t) &= \beta(t)\dfrac{S}{N}I_U  + (1 - \epsilon)\beta(t)\dfrac{S}{N}I_R +  (1 - \epsilon)\beta(t)\dfrac{SI_T}{N} \\
        &~~~~+ (1 - \epsilon)\beta_M^\alpha \dfrac{SI_M}{N} - \sigma^\alpha E\\
        _0\mathcal{D}^\alpha_{t}E_{IC}(t) &= \rho\, \epsilon \left( \beta(t)\dfrac{S}{N}I_R +  \beta(t)\dfrac{SI_T}{N} + \beta_M^\alpha \dfrac{SI_M}{N}\right) - \sigma^\alpha E_{IC}\\
         _0\mathcal{D}^\alpha_{t}E_{IF}(t) &= (1 - \rho)\epsilon \left(\beta(t)\dfrac{S}{N}I_R +   \beta(t)\dfrac{SI_T}{N} + \beta_M^\alpha \dfrac{SI_M}{N}\right) - \sigma^\alpha E_{IF}\\
       _0\mathcal{D}^\alpha_{t}I_R(t) &=\eta \sigma^\alpha E - (\gamma_{_R}^\alpha + \varphi_{_R}^\alpha) I_R\\
       _0\mathcal{D}^\alpha_{t}I_U(t) &=(1-\eta) \sigma^\alpha E - \gamma_{_U}^\alpha I_U\\
       _0\mathcal{D}^\alpha_{t}I_M(t) &=\sigma^\alpha E_{IC} - \gamma_{_M}^\alpha I_M\\
       _0\mathcal{D}^\alpha_{t}I_T(t) &=\sigma^\alpha E_{IF} - (\gamma_{_T}^\alpha + \varphi_{_T}^\alpha) I_T\\
       _0\mathcal{D}^\alpha_{t}H(t) &= \varphi_{_R}^\alpha I_{R} + \varphi_{_T}^\alpha I_{T} - (\gamma_{_H}^\alpha  + \mu_{_H}^\alpha) H\\
       _0\mathcal{D}^\alpha_{t}R(t) &= \gamma_{_R}^\alpha I_R + \gamma_{_U}^\alpha I_U + \gamma_{_M}^\alpha I_M + \gamma_{_T}^\alpha I_T +  \gamma_{_H}^\alpha H\\
       _0\mathcal{D}^\alpha_{t}C_1(t) &= \sigma^\alpha E\\
       _0\mathcal{D}^\alpha_{t}C_2(t) &= \sigma^\alpha E_{IC}\\
       _0\mathcal{D}^\alpha_{t}C_3(t) &= \sigma^\alpha E_{IF}\\
       _0\mathcal{D}^\alpha_{t}D(t) &= \mu_{_H}^\alpha H,
    \end{split}
\end{equation}
where $E_{IC}$ and $E_{IF}$ are exposed individuals who will be traced and tracked during the incubation and infectious stage, respectively. $I_M$ are infectious individuals who have been tracked while incubating and are being monitored. $I_T$ are infectious individuals who are symptomatic when tracked and will be removed or isolated. The last four equations in $(\ref{eqn:CTmodel})$ are used to estimate the cumulative total cases (both unreported and reported cases whose contacts are not being traced), cumulative cases of traced persons who will be tracked while incubating,  cumulative cases of traced persons who are infectious when tracked and the  resulting cumulative deaths from the impact of CT. We shall consider the following three special cases:
\subsubsection{Perfect Monitoring and Tracking}
In this case, we assume that the tracked and monitored contacts do not cause secondary infections, in which case $\beta_M = 0$ and that all traced contacts will be tracked while incubating, that is, $\rho = 1.$ The effective  reproduction number, $\mathcal{R}_c$ (see Appendix B), is given as   \[\mathcal{R}_c = \mathcal{R}_0\left[ \eta \dfrac{\gamma_{_U}^\alpha}{\gamma_{_R}^\alpha + \varphi_{_R}^\alpha}(1- \epsilon) + (1 - \eta)\right],\] 
where $ \mathcal{R}_0 = \beta_0^{\alpha}/\gamma_{_U}^{\alpha}$  is the {basic} reproduction number of the initial model (no contact tracing or hospitalization of cases). Thus, the contact tracing effort required to ensure that the effective reproduction number is below one is:
\[\mathcal{R}_c < 1 \Leftrightarrow \eta \left[ 1 - \dfrac{\gamma_{_{U}}^\alpha}{\gamma_{_R}^\alpha + \varphi_{_R}^\alpha}(1 - \epsilon) \right] > 1 - \dfrac{1}{\mathcal{R}_0}.\]
In the special case where we have high hospitalization rate and low recovery rates (see Table \ref{tab:Parameters}) such that $\gamma_{_U}^\alpha = \gamma_{_R}^\alpha = \varphi_{_R}^\alpha$, then
\[  0.5\eta \,(1 + \epsilon) > \left(1 - \dfrac{1}{\mathcal{R}_0}\right).\]
where $0.5\eta\,( 1 + \epsilon)$ is the critical proportion of the total cases which must be traced in order for $\mathcal{R}_c < 1$. Another special case is when we have low hospitalization rate and high recovery rates such that $\gamma_{_U}^\alpha = \gamma_{_R}^\alpha  + \varphi_{_R}^\alpha$, then

\[\eta\,\epsilon > \left(1 - \dfrac{1}{\mathcal{R}_0}\right).\]
This indicates that a larger proportion of reported cases will be traced in the former (special) case with high hospitalization and low recovery rates than the latter one with low hospitalization and high recovery rates. Now, let's rewrite $\epsilon$ as
\[\epsilon = \dfrac{\text{Number of traced contacts per reported cases}}{\text{Total number of contacts reported}}  = \dfrac{\ell}{n}, \]
and let the transmission rate $\beta_0^{\alpha}$ be written as $\beta_0^{\alpha} = p\, c^{\alpha}$, where $p$ is the probability of transmission per contact and $c$ is the contact rate. For an untraced reported case, \[ n = c^{\alpha} \left(\dfrac{1}{\gamma_{_R}^\alpha} + \dfrac{1}{\varphi_{_R}^\alpha}\right) = \beta_0^{\alpha}\dfrac{(\gamma_{_R}^\alpha + \varphi_{_R}^\alpha)}{p\,\gamma_{_R}^\alpha\,\varphi_{_R}^\alpha}.\]
Let $\kappa$ be the average number of secondary infected traced contacts identified per untraced reported case, then
\begin{equation}
    \label{eqn:k}
     \kappa := pl = \epsilon\,\beta_0^{\alpha}\dfrac{\gamma_{_R}^\alpha + \varphi_{_R}^\alpha}{\gamma_{_R}^\alpha\,\varphi_{_R}^\alpha}
\end{equation}
We note that $\kappa$ can be estimated directly from CT data and records. Also, we define the parameter $s$ as the fraction of reported cases which are traced, that is 
\begin{equation}
\label{eqn:s}
\begin{split}
    s &= \dfrac{\epsilon \left(\frac{1}{\gamma_{_R}^\alpha} + \frac{1}{\varphi_{_R}^\alpha}\right)}{\epsilon\left(\frac{1}{\gamma_{_R}^\alpha} + \frac{1}{\varphi_{_R}^\alpha}\right) + \eta(1- \epsilon)\left(\frac{1}{\gamma_{_R}^\alpha} + \frac{1}{\varphi_{_R}^\alpha}\right) + (1 - \eta)/\gamma_{_U}^\alpha}\\
    &= \dfrac{\epsilon}{\epsilon + \eta(1- \epsilon) + (1 - \eta)\frac{\gamma_{_R}^\alpha\,\varphi_{_R}^\alpha}{\gamma_{_U}^\alpha(\gamma_{_R}^\alpha+\varphi_{_R}^\alpha)}}.
\end{split}
\end{equation} 
%{From a dynamical point of view, we define the cumulative reported cases which are traced  over some time interval $[t-a, t]$ as
%\begin{equation*}
%      \tilde{s}_a(t) = \dfrac{\gamma_{_M}^\alpha \mathcal{J}\left(I_M(t) \right)}{\gamma_{_M}^\alpha \mathcal{J}_a^\alpha\left(I_M(t) \right) + (\gamma_{_R}^\alpha + \varphi_{_R}^\alpha)\mathcal{J}_a^\alpha\left(I_R(t) \right)},
% \end{equation*}
% where $\mathcal{J}_a^\alpha\left(f(t) \right)$ is the Riemann-Liouville integral of the function $f(t)$ in the interval $[t-a, t]$.
% It can be shown (see Appendix ---) that $\tilde{s}_a(t) < s$ for sufficiently large t and small $a$. See Fig. }\\ 
Using the formulas (\ref{eqn:k}) and (\ref{eqn:s}) for $\kappa$ and $s$, respectively, we obtain the formula
\begin{equation*}
    \mathcal{R}_{c} < \kappa\left( \dfrac{1-s}{s}\right) = \mathcal{R}_{c}^*,
\end{equation*}
where $\mathcal{R}_c^*$ is the product of the average number of the secondary infected traced contacts per untraced reported case and the odds that a reported case is not a traced contact.  For 100\% reporting, $s = \kappa/(\kappa+m)$ which implies that a reported case causes $\kappa+m$ secondary infections where $\kappa$ (or $m$) of these cases are traced (or untraced). Thus, $\mathcal{R}_c^* = m$
which is the fraction of secondary infected contacts to be traced that are not yet tracked. 
\subsubsection{Perfect Reporting and Tracking (Imperfect Monitoring)}
Here, we consider the case where each traced contact is tracked during the incubation stage and all infected individuals are reported. This implies that $\eta  = \rho = 1$. The reproduction number in the absence of CT  is given as \\
$\mathcal{R}_0  = \beta_0^{\alpha}/(\gamma_{_R}^\alpha + \varphi_{_R}^\alpha)$, see Appendix A. In a similar manner, the reproduction number of contact traced (monitored) person is $\mathcal{R}_M = \beta_{_M}^\alpha/\gamma_{_M}^\alpha$. Then $\theta_1 = \mathcal{R}_M/\mathcal{R}_0$ is the reduction in secondary cases of a traced (monitored) person compared to an untraced person. Thus, $\mathcal{R}_c = (1 - \epsilon)\mathcal{R}_0 + \epsilon\, \mathcal{R}_M$ and the proportion of cases to be traced so that $\mathcal{R}_c$ is below one is 
\[\epsilon > (1 - \theta_1)^{-1}\left(1 - \dfrac{1}{\mathcal{R}_0} \right). \]
Using CT observables, we describe $\mathcal{R}_c$ by defining $\kappa = \epsilon\,\mathcal{R}_0$ and $\kappa_{_M} = \epsilon\,\mathcal{R}_M$ as the average number of traced infected secondary cases per primary reported untraced and traced infected, respectively,  with $s$ given as $s = \epsilon$, then 
\begin{equation*}
    \mathcal{R}_{c} = \kappa\left( \dfrac{1-s}{s}\right)  + \kappa_{_M}.
\end{equation*}
\subsubsection{Perfect Reporting and Monitoring (Imperfect Tracking)}
Lastly, we consider perfect reporting and monitoring  with secondary traced individual during the incubation stage ( or infectious stage) with probability $\rho$ (or ($1 -\rho$)). This implies that $\beta_{M} = 0$ and $\eta = 1$. The reproduction number in the absence of CT is $\mathcal{R}_0 = \beta_0^{\alpha}/(\gamma_{_M}^\alpha + \varphi_{R}^\alpha)$ and the reproduction number of contact traced individual who are incubating or infectious when tracked is $\mathcal{R}_T = \beta_0^{\alpha}(1 - \rho)/(\gamma_{_T}^\alpha + \varphi_{_T}^\alpha)$. Thus, $\theta_2 = \mathcal{R}_T/\mathcal{R}_0$ is the reduction in secondary cases of a traced individual (who will be infectious or incubating when tracked) compared to an untraced reported case. Thus, the effective reproduction number $\mathcal{R}_c$ reduces to $\mathcal{R}_c = (1 - \epsilon)\mathcal{R}_0 + \epsilon\,\mathcal{R}_T$. As in previous cases, the critical proportion of total cases which is to be traced for $\mathcal{R}_c < 1$ is
\[ \epsilon > (1 - \theta_2)^{-1}\left( 1 - \dfrac{1}{\mathcal{R}_0}\right).\]
To describe the reproduction number in terms of CT observables, we let $\kappa_{_T} = \epsilon\,\mathcal{R}_T$ be the average number of traced infected secondary cases per primary reported traced infected with $s = \epsilon$, then 
\begin{equation*}
    \mathcal{R}_{c} = \kappa\left( \dfrac{1-s}{s}\right)  + \kappa_{_T}.
\end{equation*}

\section{Simulation Experiments and results}
\subsection{Methods and Model fitting}
We used the infected and cumulative mortality data compiled by the Center for Systems and Science Engineering at John Hopkins University (2020) \cite{JHU} starting  from the day of the first record of infection {with two intermediate days for the first 200 days (the parameters are adjusted accordingly)} in a given state to calibrate the parameter set  $(\beta_0, \kappa_1, \kappa_2,  \mu, \alpha)$ and the initial condition $E_0$. The other initial conditions are fixed, for example, $I_{R0}$ is matched with the first recorded case, $I_{U0} = (0.65/0.35)I_{R0}$ since $65\%$ of the cases are taken to be unreported and the rest are set to zero. The remaining parameters in the model are fixed at default values given in Table \ref{tab:Parameters}. Parameter fittings were performed using a nonlinear least squares algorithm in python with the limited memory BFGS method. One main benefit of the routine is the use of bounds for fit parameters. This allows faster convergence of the algorithm and ensures obtaining meaningful fit parameters. The fitted parameters and their standard errors  are given in Table \ref{tab:fittedParameters}. {A comparison of the fractional-order model with its corresponding integer-order model is given in table \ref{tab:Comparison} for California and Washington. We have excluded the states of Tennessee and Texas because their models are simply integer-order models as shown in table \ref{tab:fittedParameters} where $\alpha \approx 1$.}  All numerical simulations were done with our numerical scheme \cite{Biala2018} from which we obtain the solution of the proposed model at each time step as

\begin{enumerate}
 \item Predictor:\\ 
    \begin{equation*}
    \begin{split}
        S_p &= S_{j} + \dfrac{\tau^\alpha}{\Gamma(1 + \alpha)}F_1(t_j, S_j, E_j, I_{R, j}, I_{U, j}, {H}_j, R_j, D_j) + \tilde{H}_{1, j}\\
         E_p &= E_{j} + \dfrac{\tau^\alpha}{\Gamma(1 + \alpha)}F_2(t_j, S_j, E_j, I_{R, j}, I_{U, j}, {H}_j, R_j, D_j) + \tilde{H}_{2,j}\\
          I_{R,p} &= I_{R, j} + \dfrac{\tau^\alpha}{\Gamma(1 + \alpha)}F_3(t_j, S_j, E_j, I_{R, j}, I_{U, j}, H_j, R_j, D_j) +\tilde{H}_{3,j}\\
          I_{U,p} &= I_{U, j} + \dfrac{\tau^\alpha}{\Gamma(1 + \alpha)}F_4(t_j, S_j, E_j, I_{R, j}, I_{U, j}, H_j, R_j, D_j) + \tilde{H}_{4,j}\\
           H_p &= H_{j} + \dfrac{\tau^\alpha}{\Gamma(1 + \alpha)}F_5(t_j, S_j, E_j, I_{R, j}, I_{U, j}, H_j, R_j, D_j) + \tilde{H}_{5,j}\\
            R_p &= R_{j} + \dfrac{\tau^\alpha}{\Gamma(1 + \alpha)}F_6(t_j, S_j, E_j, I_{R, j}, I_{U, j}, H_j, R_j, D_j) + \tilde{H}_{6,j}\\
             D_p &= D_{j} + \dfrac{\tau^\alpha}{\Gamma(1 + \alpha)}F_7(t_j, S_j, E_j, I_{R, j}, I_{U, j}, H_j, R_j, D_j) + \tilde{H}_{7,j}\\
    \end{split}
\end{equation*}
    \item Corrector:\\ 
    \begin{equation*}
    \begin{split}
        S_{j+1} &= S_{j} + \dfrac{\tau^\alpha}{\Gamma(2 + \alpha)}\Big(\alpha\, F_1(t_j, S_j, E_j, I_{R, j}, I_{U, j}, H_j, R_j, D_j) \\
        & ~~~~~~~+ ~F_1(t_{j+1}, S_p, E_p, I_{R, p}, I_{U, p}, H_p, R_p, D_p)\Big) + \tilde{H}_{1, j},\\
        E_{j+1} &= E_{j} + \dfrac{\tau^\alpha}{\Gamma(2 + \alpha)}\Big(\alpha\, F_2(t_j, S_j, E_j, I_{R, j}, I_{U, j}, H_j, R_j, D_j) \\
        & ~~~~~~~+ ~F_2(t_{j+1}, S_p, E_p, I_{R, p}, I_{U, p}, H_p, R_p, D_p)\Big) + \tilde{H}_{2, j},\\
        I_{R, j+1} &= I_{R, j} + \dfrac{\tau^\alpha}{\Gamma(2 + \alpha)}\Big(\alpha\, F_3(t_j, S_j, E_j, I_{R, j}, I_{U, j}, H_j, R_j, D_j) \\
        & ~~~~~~~+ ~F_3(t_{j+1}, S_p, E_p, I_{R, p}, I_{U, p}, H_p, R_p, D_p)\Big) + \tilde{H}_{3, j},\\
        I_{U,j+1} &= I_{U,j} + \dfrac{\tau^\alpha}{\Gamma(2 + \alpha)}\Big(\alpha\, F_4(t_j, S_j, E_j, I_{R, j}, I_{U, j}, H_j, R_j, D_j) \\
        & ~~~~~~~+ ~F_4(t_{j+1}, S_p, E_p, I_{R, p}, I_{U, p}, H_p, R_p, D_p)\Big) + \tilde{H}_{4, j},\\
        H_{j+1} &= H_{j} + \dfrac{\tau^\alpha}{\Gamma(2 + \alpha)}\Big(\alpha\, F_5(t_j, S_j, E_j, I_{R, j}, I_{U, j}, H_j, R_j, D_j) \\
        & ~~~~~~~+ ~F_5(t_{j+1}, S_p, E_p, I_{R, p}, I_{U, p}, H_p, R_p, D_p)\Big) + \tilde{H}_{5, j},\\
        R_{j+1} &= R_{j} + \dfrac{\tau^\alpha}{\Gamma(2 + \alpha)}\Big(\alpha\, F_6(t_j, S_j, E_j, I_{R, j}, I_{U, j}, H_j, R_j, D_j) \\
        & ~~~~~~~+ ~F_6(t_{j+1}, S_p, E_p, I_{R, p}, I_{U, p}, H_p, R_p, D_p)\Big) + \tilde{H}_{6, j},\\
        D_{j+1} &= D_{j} + \dfrac{\tau^\alpha}{\Gamma(2 + \alpha)}\Big(\alpha\, F_7(t_j, S_j, E_j, I_{R, j}, I_{U, j}, H_j, R_j, D_j) \\
        & ~~~~~~~+ ~F_7(t_{j+1}, S_p, E_p, I_{R, p}, I_{U, p}, H_p, R_p, D_p)\Big) + \tilde{H}_{7, j},\\
    \end{split}
\end{equation*}
where 
\begin{equation*}
    \begin{split}
        F_1(t_j, S_j, E_j, I_{R, j}, I_{U, j}, H_j, R_j, D_j) &=   - \beta(t)\dfrac{S_j}{N}\left( I_{R,j} + I_{U, j}\right),\\
        F_2(t_j, S_j, E_j, I_{R, j}, I_{U, j}, H_j, R_j, D_j) &= \beta(t)\dfrac{S_j}{N}\left( I_{R,j} + I_{U, j}\right) - \sigma^\alpha E_j,\\
       F_3(t_j, S_j, E_j, I_{R, j}, I_{U, j}, H_j, R_j, D_j) &= \eta\sigma^\alpha E_j - (\gamma_{_R}^\alpha + \varphi_{_R}^\alpha)I_{R,j}, \\
        F_4(t_j, S_j, E_j, I_{R, j}, I_{U, j}, H_j, R_j, D_j) &= (1-\eta)\sigma^\alpha E_j - \gamma_{_U}^\alpha  I_{U,j},\\
        F_5(t_j, S_j, E_j, I_{R, j}, I_{U, j}, H_j, R_j, D_j) &= \varphi_{_R}^\alpha I_{R, j}  - (\gamma_{_H}^\alpha + \mu_{_H}^\alpha) H_j,\\
        F_6(t_j, S_j, E_j, I_{R, j}, I_{U, j}, H_j, R_j, D_j) &= \gamma_{_R}^\alpha I_{R,j} + \gamma_{_U}^\alpha I_{U,j}+ \gamma_{_H}^\alpha I_{H,j},\\
        F_7(t_j, S_j, E_j, I_{R, j}, I_{U, j}, H_j, R_j, D_j) &= \mu_{_H}^\alpha E_j,
    \end{split}
\end{equation*}
and 
\begin{equation*}
    \begin{split}
        \tilde{H}_{1, j} = \dfrac{\tau^\alpha}{\Gamma(2 + \alpha)}\sum_{l=0}^{j}a_{_{l, j}}\,F_1(t_l, S_l, E_l, I_{A, l}, I_{S, l}, H_l, R_l, D_l),\\
        \tilde{H}_{2, j} = \dfrac{\tau^\alpha}{\Gamma(2 + \alpha)}\sum_{l=0}^{j}a_{_{l, j}}\,F_2(t_l, S_l, E_l, I_{A, l}, I_{S, l}, H_l, R_l, D_l),\\
        \tilde{H}_{3, j} = \dfrac{\tau^\alpha}{\Gamma(2 + \alpha)}\sum_{l=0}^{j}a_{_{l, j}}\,F_3(t_l, S_l, E_l, I_{A, l}, I_{S, l}, H_l, R_l, D_l),\\
        \tilde{H}_{4, j} = \dfrac{\tau^\alpha}{\Gamma(2 + \alpha)}\sum_{l=0}^{j}a_{_{l, j}}\,F_4(t_l, S_l, E_l, I_{A, l}, I_{S, l}, H_l, R_l, D_l),\\
        \tilde{H}_{5, j} = \dfrac{\tau^\alpha}{\Gamma(2 + \alpha)}\sum_{l=0}^{j}a_{_{l, j}}\,F_5(t_l, S_l, E_l, I_{A, l}, I_{S, l}, H_l, R_l, D_l),\\
        \tilde{H}_{6, j} = \dfrac{\tau^\alpha}{\Gamma(2 + \alpha)}\sum_{l=0}^{j}a_{_{l, j}}\,F_6(t_l, S_l, E_l, I_{A, l}, I_{S, l}, H_l, R_l, D_l),\\
        \tilde{H}_{7, j} = \dfrac{\tau^\alpha}{\Gamma(2 + \alpha)}\sum_{l=0}^{j}a_{_{l, j}}\,F_7(t_l, S_l, E_l, I_{A, l}, I_{S, l}, H_l, R_l, D_l)
    \end{split}
\end{equation*}
are the memory terms of the respective population variables and 
\[ a_{_{l,j}} =  \dfrac{\tau^\alpha }{\Gamma(\alpha  + 2)}
\begin{cases}
-(j- \alpha )(j+1)^\alpha  + j^\alpha (2j - \alpha  -1 ) - (j-1)^{\alpha  +1 }, \qquad \qquad \qquad  \qquad l = 0,\\
(j-l+2)^{\alpha +1} - 3(j-l+1)^{\alpha +1} + 3(j-l)^{\alpha  +1 }- (j-l-1)^{\alpha  +1 }, \quad 1 \leq l \leq j-1,\\
2^{\alpha  +1 } - \alpha   - 3, \qquad \qquad \qquad  \qquad  \qquad \qquad \qquad \qquad  \qquad  \qquad \qquad  \qquad l = j.
\end{cases}
\]
\end{enumerate}
\begin{landscape}
\begin{table}[H]
\centering
{
\caption{Fitted Parameters to some selected States in the US, where SE denotes the standard error and CA, TN, TX and WA are acronyms for California, Tennessee, Texas and Washington, respectively. \label{tab:fittedParameters}}
\begin{tabular}{|c|c|c|c|c|c|c|c|c|c|c|c|c|}
\cline{1-13}
& \multicolumn{2}{|c|}{$E_0$} & \multicolumn{2}{|c|}{$\kappa_1$} & \multicolumn{2}{|c|}{$\kappa_2$} & \multicolumn{2}{|c|}{$\beta_0$} & \multicolumn{2}{|c|}{$\mu_{_H}$} & \multicolumn{2}{|c|}{$\alpha$}\\
\cline{1-13}
States  & Value & SE & Value & SE & Value & SE & Value & SE & Value & SE & Value & SE \\
\cline{1-13}
CA & 4522 & 211 & 0.1704 & 0.0456 & 7.6e-09 & 1.7e-10 & 0.6065 & 0.0853 & 9.8e-04 & 2.7e-05 & 0.8000 & 0.1257\\
\cline{1-13}
TN & 1000 & 20 & 0.6919 & 0.1922 & 16.2174 & 2.7270 & 1.6923 & 0.2554 & 0.0026 & 0.0002 & 0.9999 & 1.2e-04\\
\cline{1-13}
TX & 1000 & 28 & 0.4068 & 0.0876 & 14.2815 & 1.3627 & 0.8909 & 0.1280 & 0.0048 & 9.2102e-04 & 0.9999 & 1.2e-05\\
\cline{1-13}
WA & 6588 & 313 & 0.5294 & 0.1443 & 3.9e-08 & 8.1e-09 & 0.9813 & 0.3581 & 0.0013 & 3.1465e-04 & 0.8000 & 0.0965\\
\cline{1-13}
\end{tabular}
}
\end{table}
\begin{table}[H]
\centering
{
\caption{Computational Comparison of the Fractional-order model with its corresponding integer-order model. MSE denotes mean squared error, AIC and BIC denotes Akaike and Bayesian Information criterion. \label{tab:Comparison}}
\begin{tabular}{|c|c|c|c|c|c|c|}
\cline{1-7}
& \multicolumn{3}{|c|}{Integer-order model} & \multicolumn{3}{|c|}{Fractional-order model}\\
\cline{1-7}
States  & MSE & AIC & BIC & MSE & AIC & BIC \\
\cline{1-7}
CA & 4.216e+10 & 1540.58 & 1553.60 & 1.724e+10 & 1454.24 & 1469.86\\
\cline{1-7}
WA & 1.117e+09 & 1177.49 & 1190.51 & 5.555e+08 & 1110.69 & 1126.32\\
\cline{1-7}
\end{tabular}
}
\end{table}

\end{landscape}
% \begin{figure}[th]
%     \centering
%     \includegraphics[width=1.2\textwidth]{simulation}
%     \caption{Data and model fits for some selected states in the US.\label{fig:Fit}}
% \end{figure}
\begin{figure}[ht]
  \begin{subfigure}[b]{0.6\textwidth}
    \centering
    \includegraphics[width=1.0\textwidth]{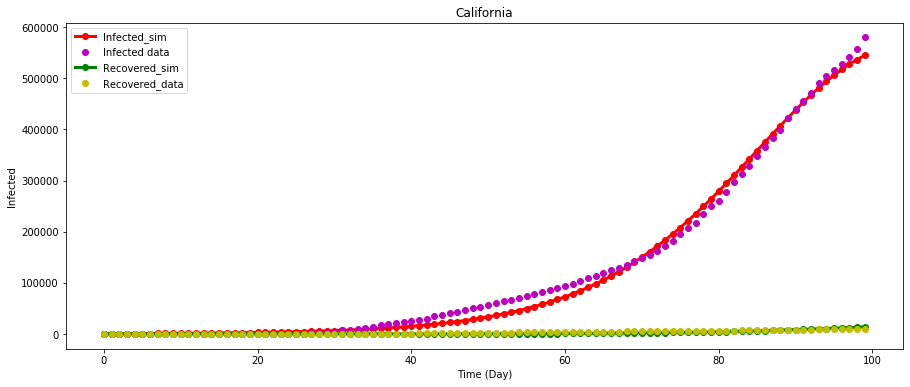} 
  \end{subfigure}%% 
  \begin{subfigure}[b]{0.6\textwidth}
    \centering
    \includegraphics[width=1.0\textwidth]{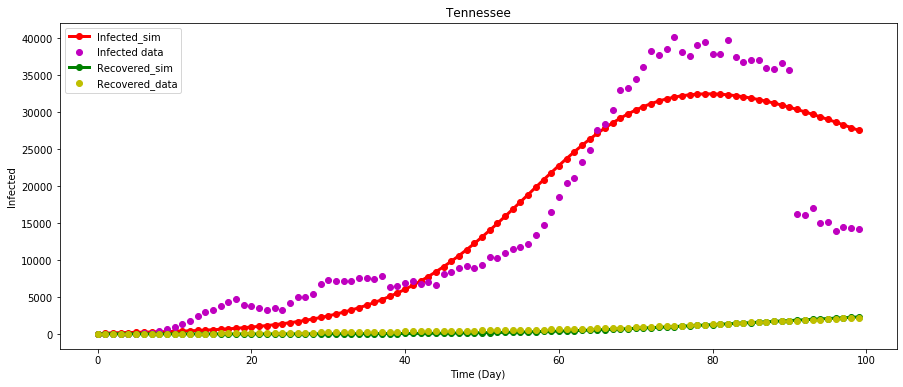}
  \end{subfigure}%% 
  \\
  \\
   \begin{subfigure}[b]{0.6\textwidth}
    \centering
    \includegraphics[width=1.0\textwidth]{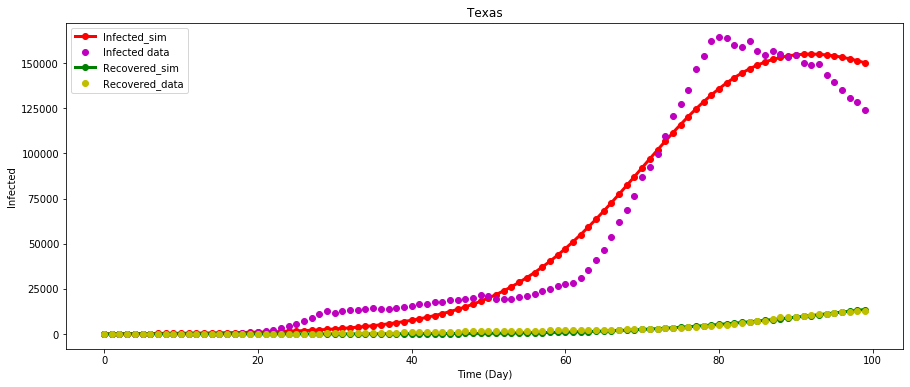} 
  \end{subfigure}%% 
  \begin{subfigure}[b]{0.6\textwidth}
    \centering
    \includegraphics[width=1.0\textwidth]{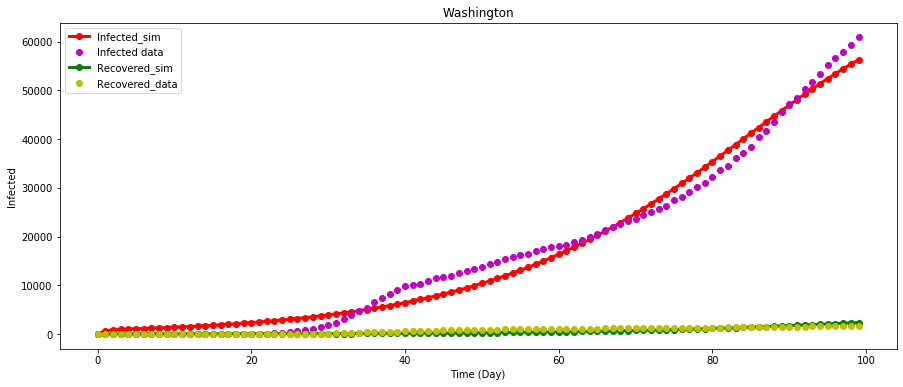}
  \end{subfigure}%% 
  \caption{Data and model fits for some selected states in the US.\label{fig:Fit}}
\end{figure}

\subsection{CT in Simulated Model with Perfect Tracking and Monitoring}
{We note that the dynamics of the model for Tennessee and Texas  are quite similar and results to an integer-order model, that $\alpha \approx 1$ (see table \ref{tab:fittedParameters} and fig.\,\ref{fig:Fit}). Thus, we consider results for Texas and Washington which have similar results with Tennessee and California, respectively in the following sections.}
\subsubsection{Immediate CT Adoption with Perfect Tracking and Monitoring}
\begin{figure}[th]
  \begin{subfigure}[b]{1.0\textwidth}
    \centering
    \includegraphics[width=1.0\textwidth]{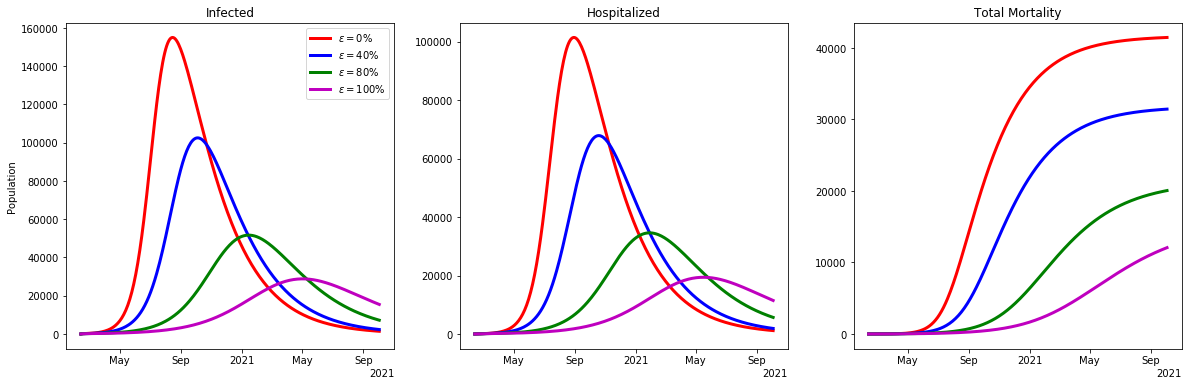} 
  \end{subfigure}%% 
  \\
  \\
  \begin{subfigure}[b]{1.0\textwidth}
    \centering
    \includegraphics[width=1.0\textwidth]{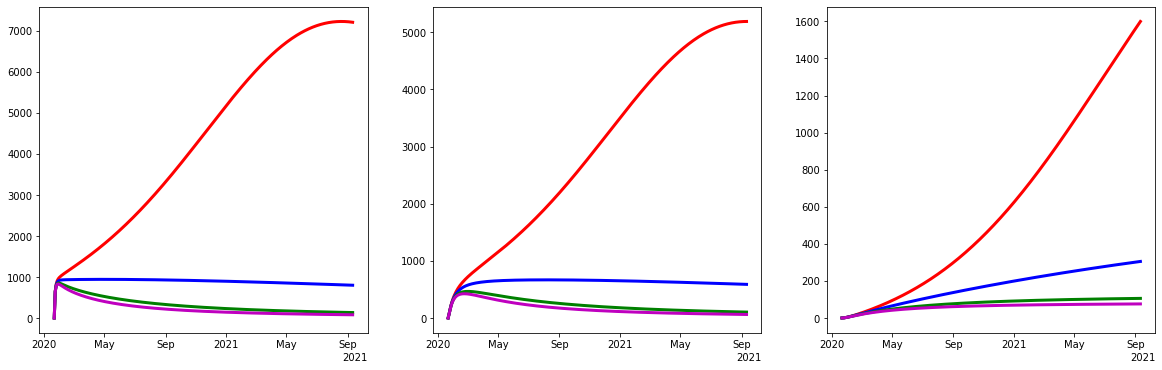}
  \end{subfigure}%% 
  \caption{Current infected, hospitalized and total mortality with varying fraction of traced reported cases in a perfect tracking and monitoring case. The first (second) row is for Texas (Washington) State. \label{fig:TMEpsilon}}
\end{figure}

\begin{figure}[ht]
  \begin{subfigure}[b]{1.0\textwidth}
    \centering
    \includegraphics[width=1.0\textwidth]{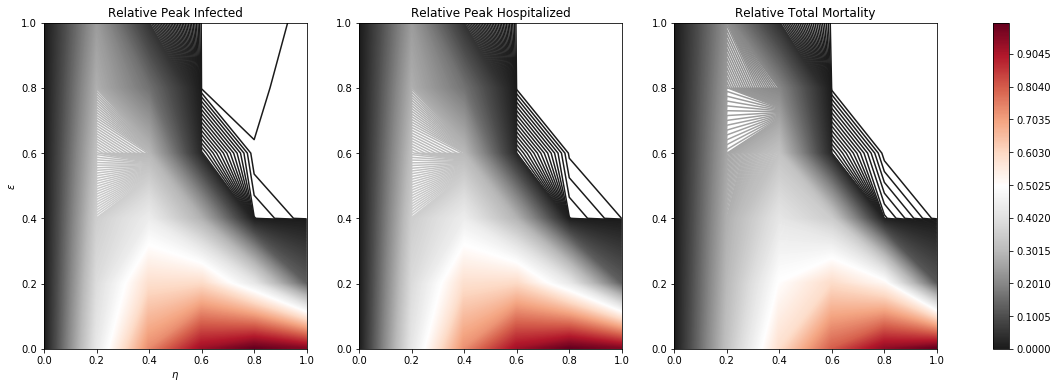} 
  \end{subfigure}%% 
  \\
  \\
  \begin{subfigure}[b]{1.0\textwidth}
    \centering
    \includegraphics[width=1.0\textwidth]{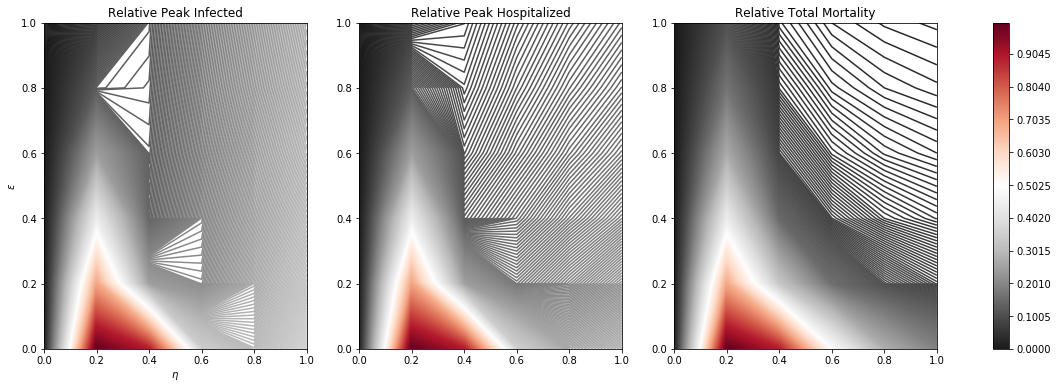}
  \end{subfigure}%% 
  \caption{Relative peak infected, hospitalizations and total mortality simulated epidemics under different reporting and tracing levels. The first (second) row is for Texas (Washington) State. \label{fig:TMContour}}
\end{figure}

\begin{figure}[ht]
    \centering
    \includegraphics[width=1.0\textwidth]{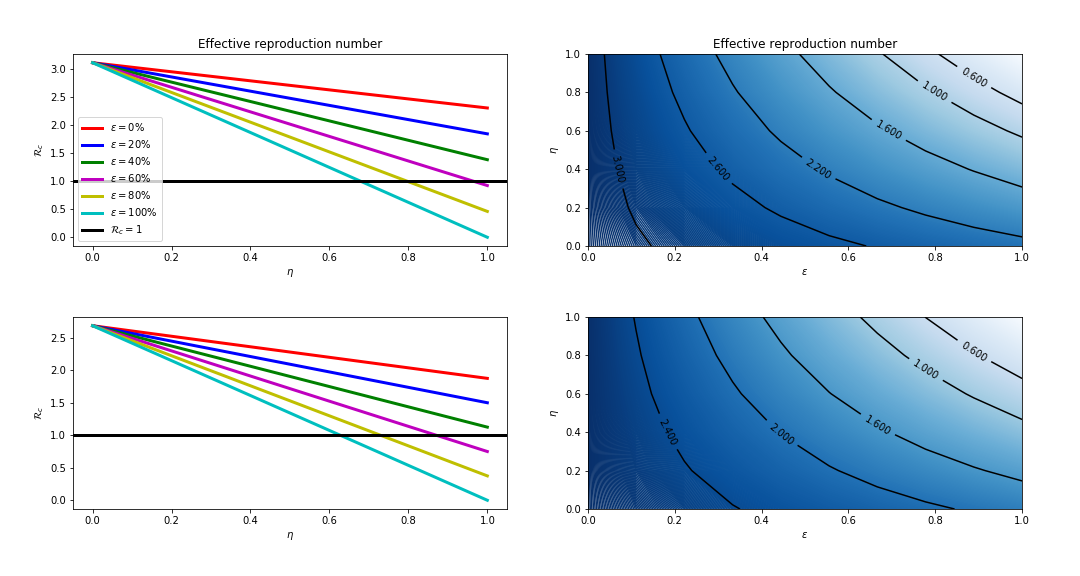} 
    \caption{Effect of CT. The first column shows profiles of the control reproduction number as a function of proportion of reported cases ($\eta$). The second column shows contour plots of the control of reproduction number as a function of proportion of reported cases ($\eta$) and traced individuals ($\epsilon$). The first (second) row is for Texas (Washington). \label{fig:R0-EpsEta}}
\end{figure}

We run the simulated model with $\beta_M = 0$ and $\rho = 1$ for around 20 months under constant conditions while studying the effect of the number of traced reported cases on the number of infected, hospitalized and dead. Fig.\,\ref{fig:TMEpsilon} shows that the total mortality (as well as infected and hospitalized) increases with no contact traced individual ($\epsilon = 0$) and decreases with increased number of traced reported cases. Furthermore, we simulate the model with several values in  $\epsilon \times \eta, ~\epsilon, \eta  \in [0, 1]$ to observe the effect of reporting and tracing on the model. The outcomes of interest are total mortality, peak hospitalization and peak infected which are normalized  against their respective maximum and the results are presented in Fig.\,\ref{fig:TMContour}. The results for both states indicate that while high reporting rate is crucial for mitigating the spread of the pandemic, the percentage of traced reported cases have a more substantial effect on the spread.
%The results in this figure show that while high reporting rate is crucial in mitigating  the spread of the epidemic, the percentage of traced reported cases have a more substantial effect on the spread. {The results for Texas are expected since the mortality and hospitalized cases will be more with more reporting and  no tracing. The results for Washington shows that a moderately high (and not very high) reporting rate gives peak hospitalizations and mortality. This may be attributed to the dynamics of the fitted model for Washington where the  number of infected individuals have been lowered after some months.}
%Using the formula given in eqn.\,\ref{eqn:s}, we estimate the number of reported cases which will be traced. The results are presented in fig.\,\ref{fig:TMTraced}.
A contour plot of the reproduction number $\mathcal{R}_c$ as a function of fraction of the infected population reported and proportion of exposed individuals that is traced is shown in fig.\,\ref{fig:R0-EpsEta}. {The figure shows that if at least $60\%~(50\%)$ contacts of reported cases (with perfect reporting)  are traced or at least $70\%~(65\%)$ of total cases are reported where all their contacts are traced in Texas (Washington), then disease elimination is feasible.}

%\begin{figure}[ht]
 %   \centering
  %  \includegraphics[width=1.0\textwidth]{EstimateTraced.png} 
   % \caption{Estimated number of reported cases traced for Texas and Washington \label{fig:TMTraced}}
%\end{figure}

\subsubsection{CT Intervention after delay}
\begin{figure}[ht]
  \begin{subfigure}[b]{1.0\textwidth}
    \centering
    \includegraphics[width=1.0\textwidth]{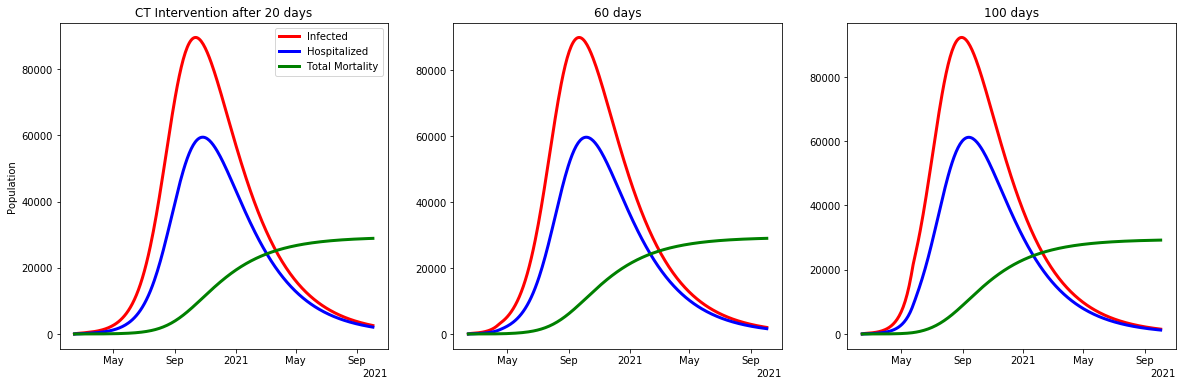} 
  \end{subfigure}%% 
  \\
  \\
  \begin{subfigure}[b]{1.0\textwidth}
    \centering
    \includegraphics[width=1.0\textwidth]{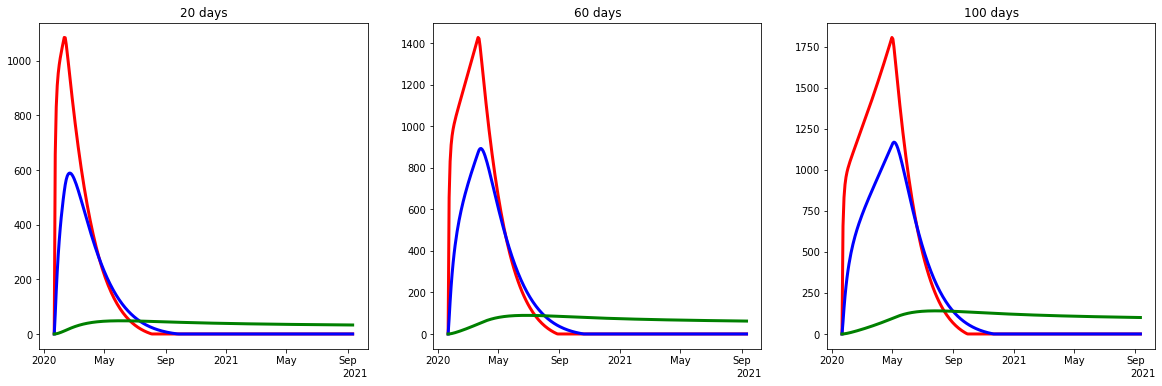}
  \end{subfigure}%% 
  \caption{CT Intervention after some discrete time delay. First (second) row is for Texas (Washington) State.  \label{fig:CTIntervention}}
  \end{figure}

We run the simulated CT model by assuming that CT was only introduced after some discrete time delay (20, 60 and 100 days). The fraction of reported cases traced is fixed at 50\%. We observe, in fig.\,\ref{fig:CTIntervention}, that the intervention of CT reduces the number of infected, hospitalization and mortality even with a late intervention time (100 days). {In Washington where the number of infected and hospitalized seems to be growing rapidly, we observe that the CT intervention drastically reduces these numbers and would be very efficient in reducing the spread of the virus in such states.}
\subsubsection{Perfect Reporting and Tracking}
\begin{figure}[ht]
  \begin{subfigure}[b]{0.95\textwidth}
    \centering
    \includegraphics[width=0.95\textwidth]{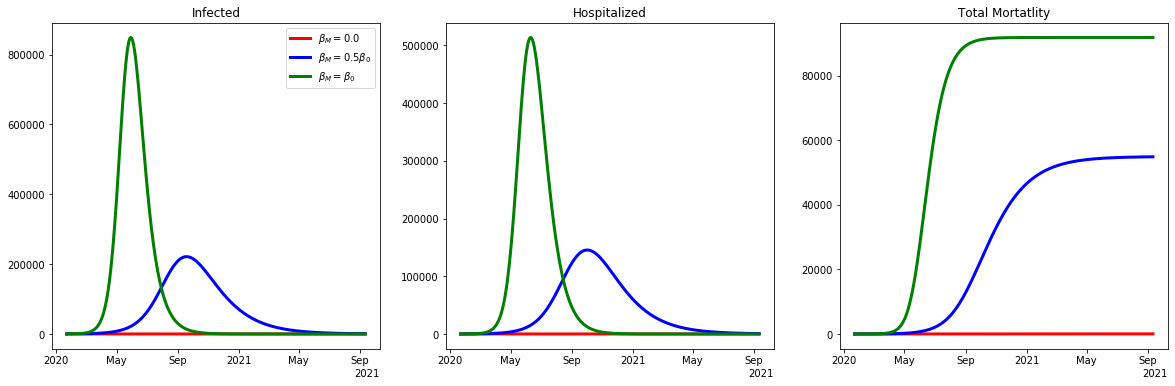} 
  \end{subfigure}%% 
  \\
  \\
  \begin{subfigure}[b]{0.95\textwidth}
    \centering
    \includegraphics[width=0.95\textwidth]{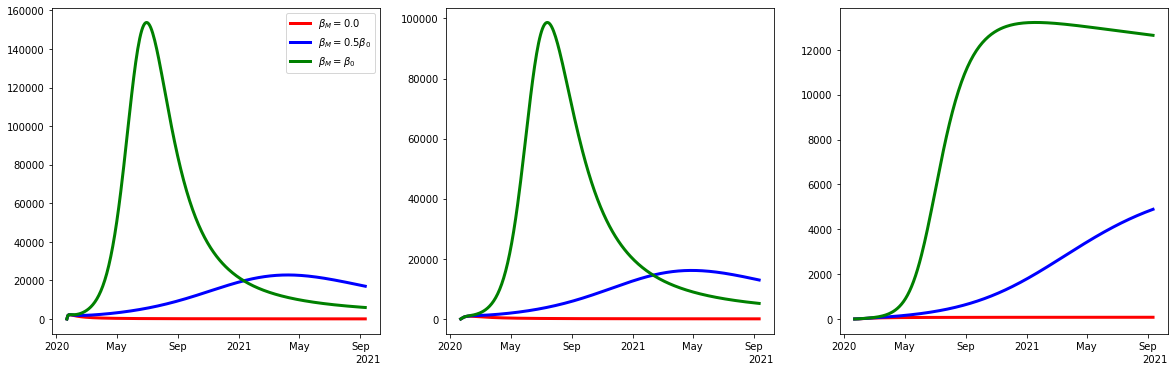}
  \end{subfigure}%% 
  \caption{Efficiency of monitoring policy in CT. The $\beta_M$ are selected to indicate 0\%, 50\% and 100\% (corresponding to $\beta_M = \beta_0, ~\beta_0/2$ and $0$, respectively) effective  monitoring policy. First (second) row is for Texas (Washington). \label{fig:RTbetaM}}
  \end{figure}

\begin{figure}[ht]
  \begin{subfigure}[b]{0.95\textwidth}
    \centering
    \includegraphics[width=0.95\textwidth]{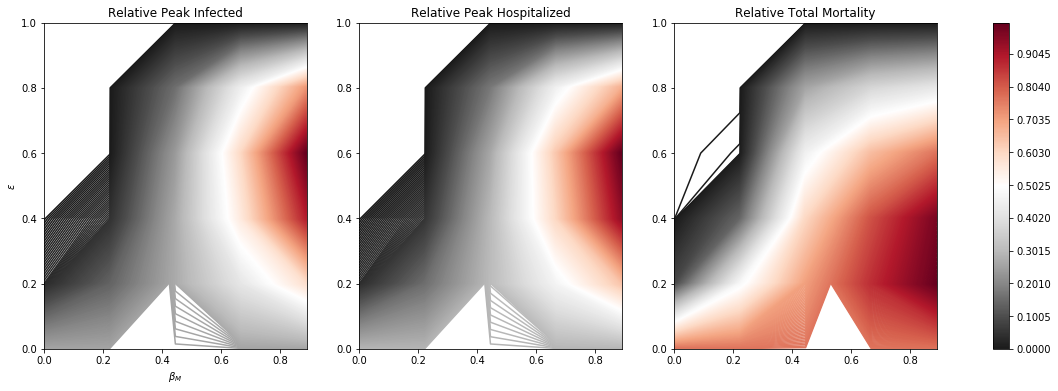} 
  \end{subfigure}%% 
  \\
  \\
  \begin{subfigure}[b]{0.95\textwidth}
    \centering
    \includegraphics[width=0.95\textwidth]{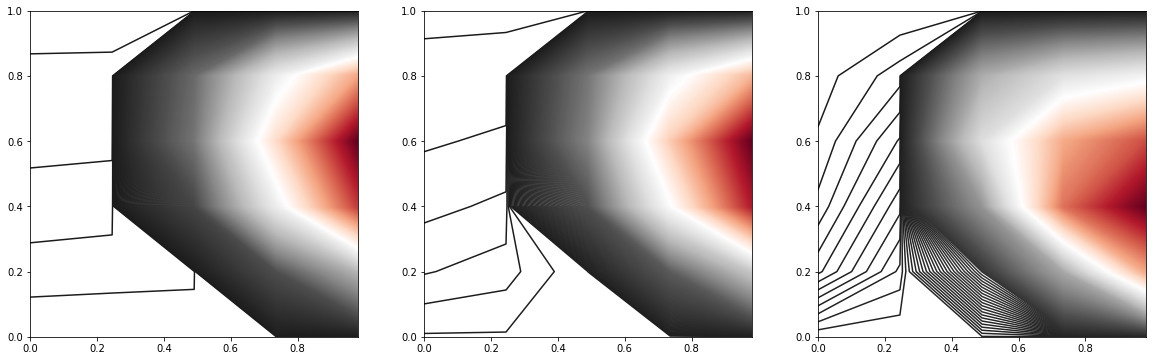}
  \end{subfigure}%% 
  \caption{Relative peak infected, hospitalizations and total mortality of simulated epidemics under different monitoring conditions and fraction of traced reported cases. First (second) row is for Texas (Washington) State. \label{fig:RTContour}}
  \end{figure}

\begin{figure}[ht]
    \centering
    \includegraphics[width=1.0\textwidth]{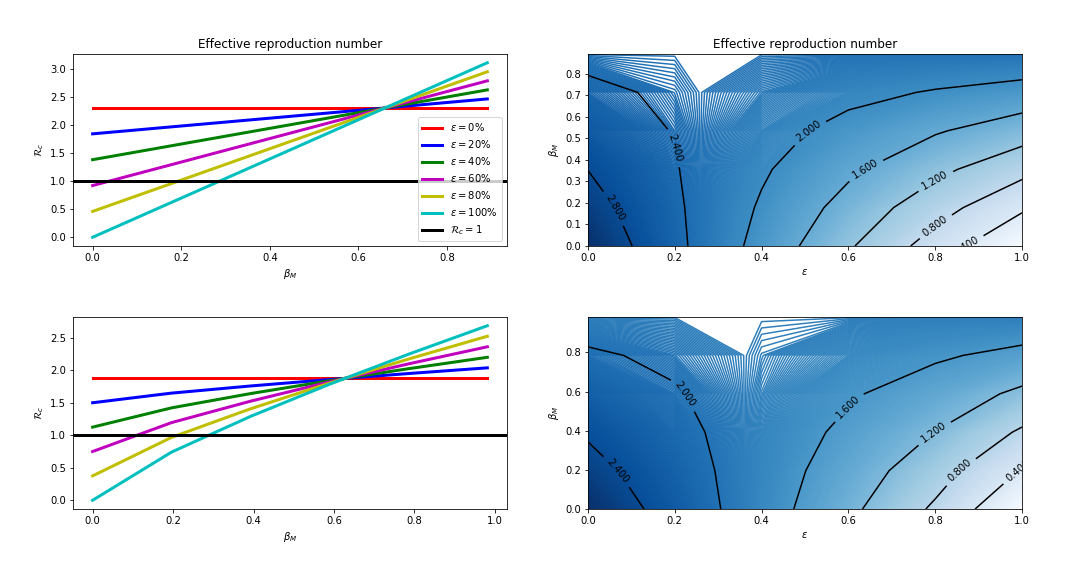} 
    \caption{Effect of CT. The first column shows profiles of the control reproduction number as a function of monitoring efficacy ($\beta_M$). The second column shows contour plots of the control of reproduction number as a function of monitoring efficacy ($\beta_M$) and proportion of traced individuals ($\epsilon$). The first (second) row is for Texas (Washington). \label{fig:R0-EpsBetaM}}
\end{figure}

We run the simulated CT model with $\rho = \eta = 1$ where CT was immediately introduced.  The fraction of reported cases traced was fixed at 50\%. We examine the effect of monitoring policy on the number of infected, hospitalizations and mortality. Fig.\,\ref{fig:RTbetaM} shows that 50\% effective monitoring policy reduces the hospitalization and total mortality.  Furthermore, we run several simulations with values in $\epsilon \times \beta_M, ~\epsilon, ~\beta_M \in [0, 1]$ and the results are shown in fig.\,\ref{fig:RTContour}. Similar to previous contour plots, the outcome of interests are relative peak hospitalization and total mortality. The observation is quite revealing: the peak hospitalizations and cumulative mortality occur when $\beta_M \approx \beta_0$ and the fraction of traced reported cases is around 20-80\%. This shows that, in general, the CT should be followed up with effective monitoring policy. A contour plot of the reproduction number $\mathcal{R}_c$ as a function of the monitoring efficacy and proportion of exposed individuals that is traced is shown in fig.\,\ref{fig:R0-EpsBetaM}. The figure shows that the disease will die out if all the traced individuals are being monitored so that they are about $70\%~(75\%)$ noninfectious as an unreported individuals or untraced infectious cases or with at least $60\%~(60\%)$ of reported cases being traced with a perfect monitoring policy in  Texas (Washington).

\subsubsection{Perfect Reporting and Monitoring}
\begin{figure}[ht]
  \begin{subfigure}[b]{0.95\textwidth}
    \centering
    \includegraphics[width=0.95\textwidth]{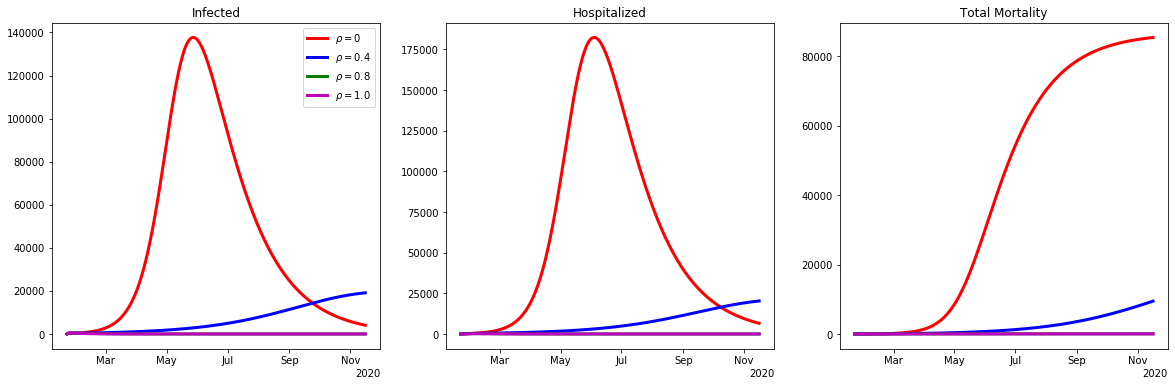} 
  \end{subfigure}%% 
  \\
  \\
  \begin{subfigure}[b]{.95\textwidth}
    \centering
    \includegraphics[width=0.95\textwidth]{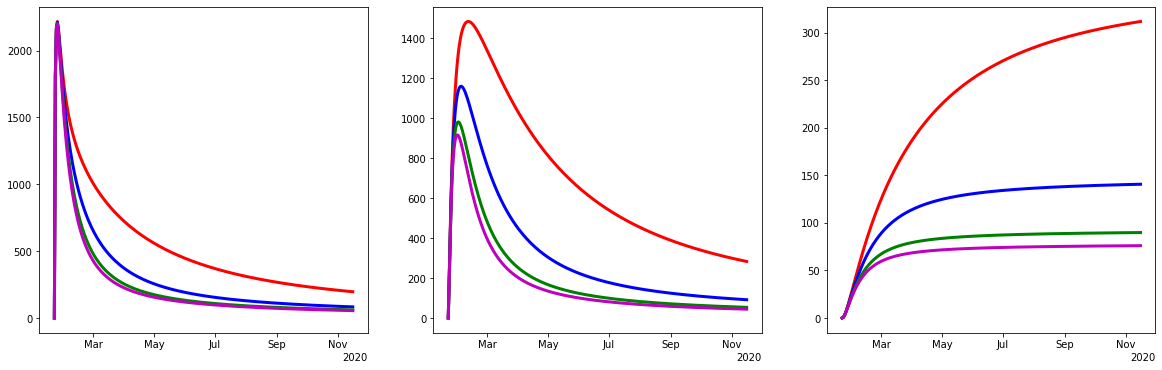}
  \end{subfigure}%% 
  \caption{Effects of tracking contacts of reported cases when incubating or being infectious. The $\rho$ values are selected to show 0\%, 40\%, 80\% and 100\% of traced reported cases are incubating when tracked. Perfect tracking implies $\rho = 1$. First (second) row is for Texas (Washington) State. \label{fig:RMrhoV}}
  \end{figure}

\begin{figure}[ht]
  \begin{subfigure}[b]{0.95\textwidth}
    \centering
    \includegraphics[width=0.95\textwidth]{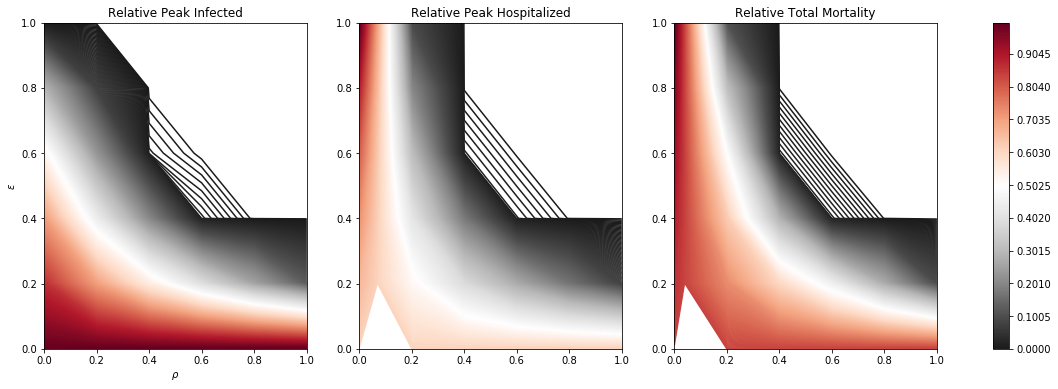} 
  \end{subfigure}%% 
  \\
  \\
  \begin{subfigure}[b]{0.95\textwidth}
    \centering
    \includegraphics[width=0.95\textwidth]{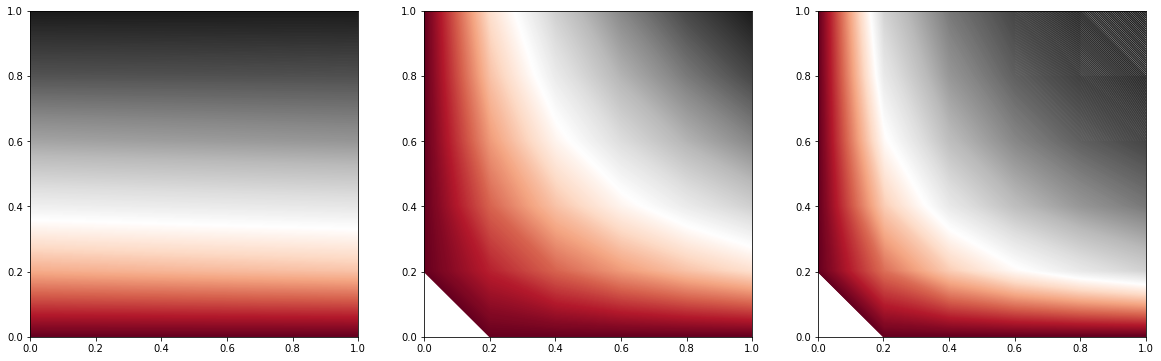}
  \end{subfigure}%% 
  \caption{Relative peak infected, hospitalizations and total mortality of simulated epidemics under different monitoring conditions and fraction of traced reported cases. First (second) row is for Texas (Washington) State. \label{fig:RMContour}}
  \end{figure}

\begin{figure}[ht]
    \centering
    \includegraphics[width=1.0\textwidth]{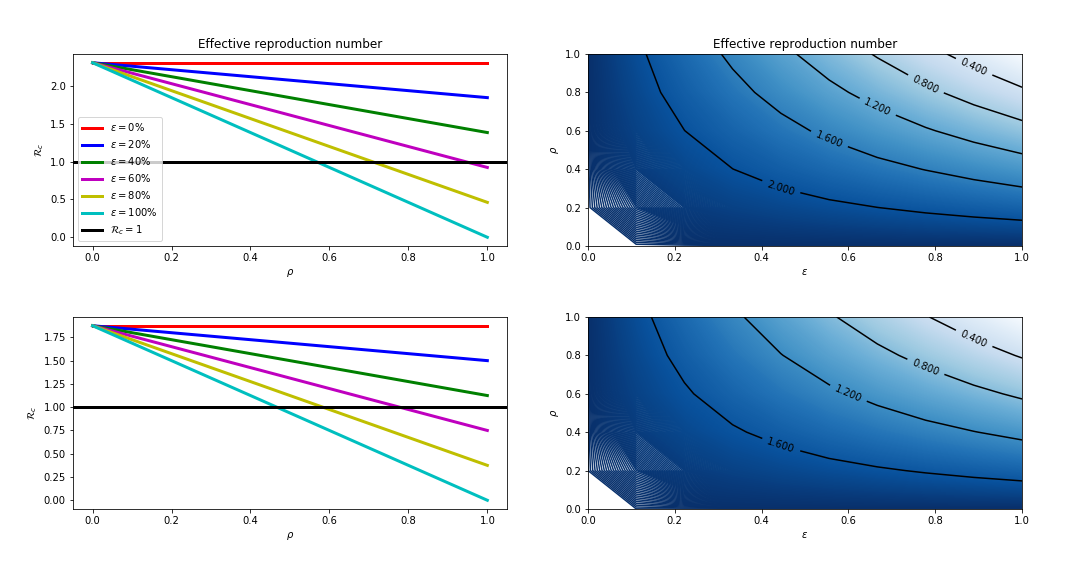} 
    \caption{Effect of CT. The first column shows profiles of the control reproduction number as a function of tracking efficacy ($\rho$). The second column shows contour plots of the control of reproduction number as a function of tracking efficacy ($\rho$) and proportion of traced individuals ($\epsilon$). The first (second) row is for Texas (Washington). \label{fig:R0-EpsRho}}
\end{figure}

In this case, we consider the numerical experiment where we assume that every infected case is reported ($\eta = 1$) and tracked contacts of reported cases are effectively monitored ($\beta_M = 0$) so that they do not cause secondary infections. We run the simulated CT model  under constant conditions with the aim of exploring the effect of $\rho$ (the fraction or probability that a traced reported case is incubating when tracked) on peak hospitalization and mortality. Unsurprisingly, we see that the higher the fraction of tracked contacts who are incubating the lower the number of hospitalizations and deaths. These results are evident in figures \ref{fig:RMrhoV}--\ref{fig:R0-EpsRho}.
  
  \clearpage
\section{Discussions and Conclusions}
The novel coronavirus epidemic which began in China has spread across the globe with over 2,000,000 deaths. Several control measures have been taken by health and government officials to mitigate the spread of the virus. Such measures include social distancing, use of face-masks, {lockdowns} and contact tracing. {Although, CT is known to be efficient in mitigating the spread of disease outbreaks, we focus on quantifying this efficacy and showing the actions of contact tracers that are mostly required in stopping the spread of the COVID-19 epidemic.} In particular, we consider special cases where we have perfect tracking and monitoring, perfect reporting and tracking, and perfect reporting and monitoring.\\
We have developed a time-fractional order differential equation  model of the contact tracing process in the COVID-19 outbreak. Our deterministic model links the action of contact tracers such as monitoring and tracking to the number of reported cases traced. Our framework separates the infected population into unreported and reported, and further splitting the reported cases into fraction whose contacts will be traced. Additionally, we incorporate the effect of tracking by considering the probability that a traced contact will be incubating (or infectious) when tracked. This inherent structure in the model captures the dynamics of contact tracing and enables us to express the reproduction number in terms of observable quantities. In particular, under the assumption that there is a perfect tracking and monitoring, we gave an upper bound for the effective reproduction number as $\mathcal{R}_c < \kappa(1-s)/s$, where $\kappa$ is the average number of secondary infected individuals traced per reported untraced case  and $(1-s)/s$ is the odds that a reported case is not a traced contact. In the case of perfect tracking with either perfect monitoring or perfect reporting, we obtain the result $\mathcal{R}_c = \kappa(1-s)/s + \kappa_{_M}$, where $\kappa_{_M}$ is the average number of secondary infected individuals per reported traced case. With these observable quantities, these formulas can provide a quick and simple estimates of the reproduction number in the population. Furthermore, we estimated the proportion of contacts that need be traced to ensure that the reproduction number is below one. Although, we would have loved to provide daily or weekly estimates of  $\mathcal{R}_c$ from the formulas (above) involving observable quantities but we were unable to find CT data for the COVID-19 epidemic. However, we relied on model simulations to gain insights on the impact of CT with different special cases and during different stages of the epidemic. In fact, the decline of peak hospitalizations and total deaths in the simulations of CT model compared to the preliminary model shows  its efficacy.  \\
With the simulated CT model, the efficiency of CT in mitigating the spread of the virus and altering the epidemiological outcomes of peak hospitalizations and total deaths is a nonlinear function of the fraction of infected cases reported, the monitoring policy and the proportion of traced contacts who are tracked while incubating (see fig.\,\ref{fig:TMContour}, \ref{fig:RTContour} and \ref{fig:RMContour}). {In the first case (``perfect tracking and monitoring") and considering that 35\% of infected cases are reported  with 40\%, 80\% and 100\% of reported cases being traced, the peak hospitalizations are reduced by 33.1\% (87.1\%), 65.9\% (91.0\%), and 80.9\% (91.8\%), respectively, for Texas (Washington) state. The total mortality is also seen to decline by 24.1\% (80.9\%), 51.6\% (93.4\%) and 70.9\% (95.3\%) in Texas (Washington). Furthermore, we investigated the intervention of CT after some discrete time delay. We observe that early intervention of CT may greatly reduce peak hospitalizations and total mortality. Even with a late intervention (after 100 days), we see that the total mortality is reduced by a factor of 39.0\% (91.2\%) in Texas (Washington).}\\
In the second case, we assumed a perfect reporting of infected cases and perfect tracking of contacts of reported cases. With 50\% of these cases traced and the monitoring policies being implemented at 50\% and 100\% efficiency, we observe the reduction in total mortality (peak hospitalizations) by 88.9\% (88.4\%) and 99.8\% (99.9\%) in Texas.  In Washington, we observed a 21.8\% (54.8\%) reduction with a 50\% efficient monitoring policy. Furthermore, the contour plots (see fig.\,\ref{fig:RTContour}) show that while both fraction of traced reported cases and the monitoring strategy are crucial in mitigating the spread, the monitoring strategy or policy is of substantial importance so that tracked reported individuals do not cause secondary infections while being monitored. Similar results are observed in the case of perfect reporting and monitoring. Finally, we showed the effects of the proportion of traced cases ($\epsilon$), monitoring efficacy ($\beta_M$), and tracking efficacy ($\rho$) in lowering the reproduction number so that the disease eventually die out after a period of time. \\
In conclusion, our findings suggests that almost all states in the US should adopt (if not yet) CT measures. In particular, our findings show that tracking a larger proportion of traced contacts while incubating and perfect monitoring of tracked contacts so that they do not cause secondary infections are  highly important for the impact of CT to be seen. \\
\indent {There are some limitations to the model discussed in this manuscript. For a new epidemic like COVID-19, we expect and have seen that limited resources have hindered governmental actions in mitigating the spread of the disease. Thus, some of the model assumptions  such as ``perfect tracking or reporting or monitoring'' are simply for mathematical quantifications. Also, we have assumed that the  probability  for fraction of first or higher order traced contacts who will be incubating and infectious, respectively, when tracked are the same. This, in general, may not be the situation.}

{\section*{Acknowledgement}The authors are grateful to the anonymous reviewers for their comments and suggestions which have greatly improved the quality of the paper.}
\section*{Appendix A. Effective Reproduction Number of Preliminary Model}
The basic reproduction number $\mathcal{R}_c$ can be defined using the next generation matrix \cite{Diekmann1990, VanDriessche2002, Diekmann2009}. The disease-free equilibrium point of the system is $\varepsilon_0 = (S_0, 0, 0, 0 )$. We define a next-generation matrix by considering the linearized system at the disease-free equilibrium point $\varepsilon_0$. By using the notations in \cite{Diekmann1990, VanDriessche2002}, it follows that the matrices $\mathscr{F}$ of new infection terms  and $\mathscr{V}$ of transfer of infection to and from the compartments are given, respectively, as
\begin{equation*}
\mathscr{F} = \left[
    \begin{array}{ccc}
     0 &  {\beta_0^\alpha} &  {\beta_0^\alpha}\\
     0 & 0 & 0 \\
     0 & 0 & 0
\end{array}
\right], ~~~
\mathscr{V} = \left[
    \begin{array}{ccc}
     \sigma^\alpha  & 0 & 0   \\
     -\eta \sigma^\alpha & \gamma_{_R}^\alpha + \varphi_{_R}^\alpha & 0\\
     - (1 - \eta)\sigma^\alpha & 0 & \gamma_{_U}^\alpha
\end{array}
\right],
\end{equation*}
{where $\beta_0^\alpha = \max \beta(t)$ is used to estimate the transmission rate.}  The effective reproduction number of the model, denoted by $\mathcal{R}_0$, is given by {\[\mathcal{R}_0 = \beta_0^\alpha\left(\dfrac{\eta}{\gamma_{_R}^\alpha + \varphi_{_R}^\alpha} + \dfrac{1 - \eta}{\gamma_{_U}^\alpha}\right).\]}
\section*{Appendix B. Effective Reproduction Number of  Model with CT}
In a similar manner to the results in Appendix A, the matrix $\mathscr{F}$ of new infections and $\mathscr{V}$ of transfer terms are given by
\begin{equation*}
\mathscr{F} = \left[
    \begin{array}{ccccccc}
     0 &  0 & 0 & (1 - \epsilon){\beta_0^\alpha} &  {\beta_0^\alpha} & (1 - \epsilon)\beta_{_M}^\alpha & \epsilon\,{\beta_0^\alpha}\\
      0 & 0 & 0 & \rho\,\epsilon\, {\beta_0^\alpha}  & 0 & \rho\,\epsilon\, \beta_{_M}^\alpha & \rho\,\epsilon\,{\beta_0^\alpha}\\
       0 & 0 & 0 & (1-\rho)\epsilon\, {\beta_0^\alpha}  & 0 & (1-\rho)\epsilon\, \beta_{_M}^\alpha & (1 -\rho)\epsilon\,{\beta_0^\alpha}\\
       0 & 0 & 0 & 0 & 0 & 0 & 0\\
       0 & 0 & 0 & 0 & 0 & 0 & 0\\
       0 & 0 & 0 & 0 & 0 & 0 & 0\\
       0 & 0 & 0 & 0 & 0 & 0 & 0
\end{array}
\right]
\end{equation*}
\begin{equation*}
\mathscr{V} = \left[
    \begin{array}{ccccccc}
     \sigma^\alpha  & 0 & 0  & 0 & 0 & 0 & 0  \\
      0 & \sigma^\alpha  & 0 & 0  & 0 & 0 & 0   \\
       0 & 0 & \sigma^\alpha  & 0 & 0  & 0 & 0 \\
     -\eta \sigma^\alpha & 0 & 0 & \gamma_{_R}^\alpha + \varphi_{_R}^\alpha & 0 & 0 & 0\\
     - (1 - \eta)\sigma^\alpha & 0 & 0 & 0 &  \gamma_{_U}^\alpha & 0 & 0\\
     0 & -\sigma^\alpha & 0 & 0 & 0 & \gamma_{_M}^\alpha & 0\\
     0 & 0 & -\sigma^\alpha & 0 & 0 & 0 & (\gamma_{_T}^\alpha + \varphi_{_T}^\alpha)
\end{array}
\right].
\end{equation*}
The effective reproduction number cannot be written explicitly here. However, the given matrices are used to obtain the reproduction numbers for each of the special cases given in the text.
 \bibliography{main}
\end{document}